# A Bioinspired Aquatic Machine Mimicking Water Caltrop

Yuanquan Liu[1], Thomas Speck[1], and Isabella Fiorello[1(✉)]

[1]Cluster of Excellence livMatS @ FIT – Freiburg Center for Interactive Materials and Bioinspired Technologies, University of Freiburg, D-79110 Freiburg, Germany
`isabella.fiorello@livmats.uni-freiburg.de`

**Abstract.** Plants are increasingly becoming a source of inspiration for robotics and engineers to develop bioinspired, adaptive, and multifunctional machines. In this study, we propose a bioinspired aquatic machine that mimics the fruit of the water caltrop (*Trapa natans* L.). Among various plant species, *T. natans* produces unique woody fruits that can disperse passively via water currents or by clinging to boats or waterfowls. Inspired by the structures and dispersal mechanisms of *T. natans*, we designed miniaturized biomimetic machines capable of passive dispersion in aquatic ecosystems. In order to study our selected biological model, we collected natural fresh and dried mature samples of *T. natans* fruits. We designed biomimetic aquatic machines by extracting the main geometrical details from the natural samples, and by exploiting advanced three-dimensional reconstruction techniques, including x-ray micro-computed topography (Micro-CT). Then, we successfully fabricate the biomimetic machines at high-resolution in two configurations (hollow body and solid body) using light-based bioprinting of photo-responsive hydrogels. We also characterized the mechanical properties of the bioprinted materials through compression tests. Finally, we evaluated the floating behavior of the biomimetic machines in a flow chamber as a *proof of concept*. This biomimetic approach enhances the adaptability of the machine in aquatic environments, offering new design insights for underwater, soft, and microrobotics.

## 1 Introduction

Bioinspired soft robotics is an innovative field that translates essential biological principles into design guidelines for creating a new generation of soft robots[1]. These robots are capable of adapting to and interacting with unpredictable and complex environments and can be used for ecosystem exploration, monitoring, and conservation[2]. Leveraging principles from both plant and animal biology has led to the development of novel robotic systems capable of autonomous movement, self-actuation, and environmental adaptation.

Among living organisms, plants have evolved efficient adaptations to nearly every biome, serving as an inspiration for scientists to develop ingenious bioinspired machines[3–5]. Recently, plant-inspired and/or plant-hybrid self-dispersing robots mimicking plant seeds and fruits have been developed for targeted cargo delivery or sensing, opening different applications in precision agriculture and reforestation fields[6–11]. For example, the authors have developed a miniaturized biohybrid machine inspired by wild oat fruits able to autonomously explore the soil environment by using sister awns as biological motor, enabling precise delivery of seeds and fertilizers[12]. Although a few biomimetic robots inspired by plant seed dispersal mechanisms, such as wind dispersal and/or hygroscopically active dispersal units, have been reported, studies on bioinspired machines mimicking aquatic fruits and their dispersal mechanisms are still rare.

Beyond plant-inspired designs, aquatic robotics has also drawn inspiration from the locomotion strategies of marine animals, such as jellyfish, fish, shellfish, and cephalopods, to develop adaptive, energy-efficient propulsion systems[13–20]. These animal-inspired soft robots utilize biomimetic actuators, flexible structures, and compliant materials to navigate complex aquatic environments with minimal energy consumption, offering promising solutions for underwater exploration, ecological monitoring, and environmental remediation[20].

Among aquatic plants, the water caltrop *Trapa natans* L. is a globally distributed invasive floating plant typically found in lakes, ponds and rivers, which produces unique four-horned woody fruits with barbed spines[21] (Fig. 1). These fruits can disperse passively, being carried by water currents [22], or the spines of the fruits can act as a hitchhiker clinging to waterfowl or boats[23], before they anchor themselves to the bottom of the lake. Taking inspiration from the unique morphology, biomechanics, and dispersal mechanism of the fruits of *T. natans* can lead to novel miniaturized machines able to be passively dispersed by water flow or anchoring to unstructured surfaces for targeted deployment.

Bioprinting and 3D printing technologies have been widely employed in the fabrication of soft robots, enabling precise replication of biological structures[24]. Among bioprinting methods, extrusion-based technology is the most common for creating hydrogel-based structures[25]. Alternatively, digital light processing (DLP)-based bioprinting uses light-induced photopolymerization to produce complex, high-resolution 3D structures, facilitating the rapid prototyping of bioinspired designs with tailored mechanical properties[26, 27]. These reasons make DLP-based bioprinting technology particularly suitable for the fabrication of soft robots with complex micro-scale structures.

Here, we propose a bioprinted aquatic machine inspired by the unique structure and dispersal mechanism of the water caltrop (*T. natans*). First, we collected fresh and dried *T. natans* fruits and analyzed their structural characteristics. By extracting key geometrical details from the natural samples, we designed biomimetic aquatic machines using advanced three-dimensional reconstruction techniques, including x-ray micro-computed tomography (Micro-CT). We then fabricated the biomimetic machines at high resolution in two configurations (i.e., hollow body and solid body) using light-based bioprinting of photo-responsive hydrogels. To assess the functional properties of these biomimetic machines, we characterized the mechanical behavior of the bioprinted materials through compression tests and evaluated their floating dynamics in a flow chamber as a *proof of concept*. Compared with traditional soft aquatic robots, our machine is able to achieve passive floating and anchoring functions through its design, breaking the limitation of existing underwater soft robots that rely on power sources. This innovation offers new insights for the design of underwater, soft, and micro-robotic applications, while also opening possibilities for agricultural seeding and environmental detection in underwater environments.

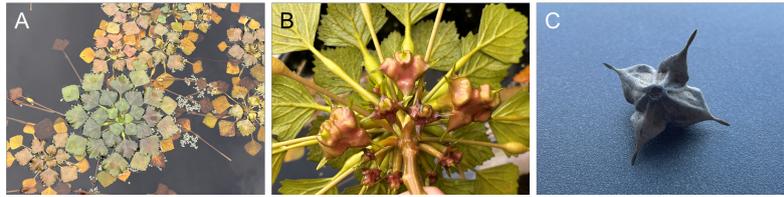

Fig. 1. (A) The water caltrop (*Trapa natans* L.) in a water tank. (B, C) Natural fruits of *T. natans*, (B) Fresh samples; (C) Mature dried sample.

## 2 Materials and Methods

### 2.1 X-Ray Micro-Computed Topography (Micro-CT) and 3D reconstruction

An X-ray micro-computed tomography system (ProCon X-Ray, PXR GmbH, Germany) was used to scan the natural dried fruits of the water caltrop (*Trapa natans* L.). The parameters were set to an accelerating voltage of 80 kV, a source current of 70 μA, an exposure time of 0.15 s, a voxel size of 8 μm, and a single scanning duration of about 20 min. After scanning, the data were reconstructed using VG-Studio Max software, and 2D slice images were exported in TIFF format. The 2D slice images were imported into Avizo software (Thermo Scientific, USA; version 9.2) for 3D reconstruction. The hollow and solid 3D models of T. *natans* were generated through segmentation, surface reconstruction, volume rendering, and filling processes, and exported as STL files. Subsequently, the STL files were imported into Meshlab and Blender software for optimization and cropping, and the final 3D models were exported as STL files for further fabrication and analysis.

### 2.2 Morphological characterizations

In this study, the morphological characteristics of natural *T. natans* and biomimetic aquatic machines were quantitatively analyzed. Natural dried *T. natans* samples were provided by the Institute of Biogeography, University of Natural Resources and Life Sciences (Vienna, Austria), while fresh *T. natans* samples were collected from the Botanical Garden of the University of Freiburg (Freiburg, Germany). The geometrical parameters (such as height, lengths, and diameters) of both natural *T. natans* and biomimetic aquatic machines were extracted using ImageJ software and a caliber. In addition, the weights of the samples were measured using a scale (KERN, EMS 300-3). An optical microscope (Nikon, SMZ25) was used to observe the microstructures of both the natural and biomimetic samples.

### 2.3  Bioprinting of machine body

We microfabricate the biomimetic machines at high resolution in two configurations: a hollow body (i.e., with an empty central space) and a solid body (i.e., fully filled with material). The machines were printed using a biocompatible photosensitive hydrogel, Polyethylene Glycol Diacrylate bioink (PEGDA X, Cellink), and a light-based DLP bioprinter (LumenX, Cellink). Prior to printing, the hollow and solid *T. natans* 3D CAD model STL files were imported into PrusaSlicer software (Prusa Research, Czech Republic) for slicing and G-code conversion, then exported in STL file format. After that, the 3D CAD models were processed into DNA Studio Illuminate software (Cellink, Sweden). A square substrate with a diameter of 1.5 mm and a height of 0.8 mm was added to the *T. natans* 3D cad model in DNA Studio Illuminate software to guarantee the adhesion between the substrate and the built platform. The biomimetic *T. natans* was bioprinted by irradiating a UV source ($\lambda$ = 405 nm) in a DLP printer, with a set printing layer height of 50μm, an exposure time of 3s, and a laser intensity of 70 %. To further improve the adhesion between the substrate and the built platform, we set the exposure time of the first three layers to 9s with a laser intensity of 100 %. Immediately after printing, we removed the substrate of the bioinspired aquatic machine sample and transferred the sample to deionized water, which was renewed several times until the water solution was clarified, thus completing the development of the sample. In addition, we used dustless paper to wipe off the solution residue on the surface of the sample and measure its weight immediately after the sample was printed. Finally, the bioprinted samples were conserved in deionized water at room temperature for further analysis and testing purposes. Additionally, we printed a 5 mm cube for mechanical testing with the same materials and printing parameters used for the biomimetic machines.

### 2.4  Mechanical characterizations under compressive loading

To characterize the mechanical properties of the bioprinted materials, we performed compressive stress-strain tests using an Inspekt Table 5 Universal tensile Testing machine (Hegewald & Peschke Meß- und Prüftechnik GmbH, Nossen, Germany) with a 500 N force transducer on a 5 mm cube (Fig. 2A). We extracted different mechanical parameters, including the Young's modulus, compressive breaking strain and compressive breaking strength. All tests on the cubes were performed at a strain rate of 0.1 mm/s. The Young's modulus was extracted from the linear region of the stress-strain curve. We also performed compressive tests on the bioprinted machines (Fig. 2B) at a strain rate of 0.1 mm/s, setting the compressive strain at 5 %. From the bioprinted machine tests, we plotted the force-displacement curves.

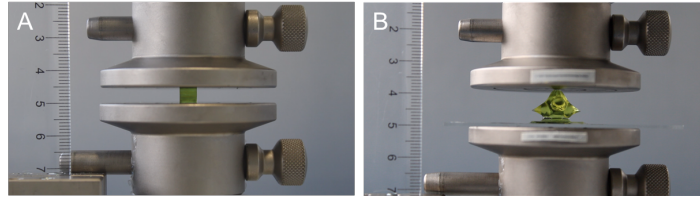

Fig. 2. Setup for compressive loading tests: (A) Cube sample, (B) Biomimetic sample (solid body).

### 2.5 Flow chamber experiments

To preliminarily investigate the motility of the bioprinted biomimetic machines (hollow body and solid body) in water, we used a custom-made flow chamber. The flow chamber consisted of Main Flow Chamber, Pump System (BADU 42/12, 230 V, Germany), Control Unit (BADU Eco Drive II 2,20 kW, Germany), Water Circulation System and Drainage System (Fig. 3). During the test, the potentiometer was set to 0, which correspond to the lowest flow rate of the machine.

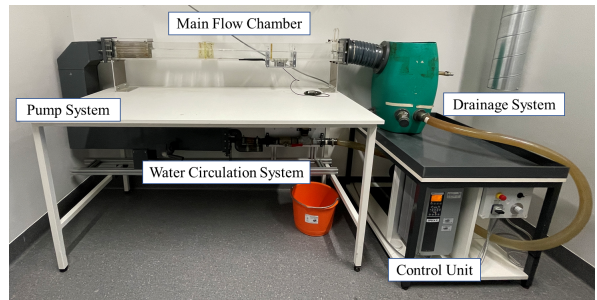

Fig. 3. Setup for flow chamber.

## 3  Results and Discussion

### 3.1  Biomimetic design

We designed bioinspired aquatic machines mimicking the morphological structures of the fruits of the water caltrop (*T. natans* L.). Firstly, we analyzed the morphology of natural fresh and dried mature fruits of *T. natans*. We observed that the pericarp of *T. natans* fruits has four corners with sharp spines attached (Fig. 1C) [21]. The pericarp undergoes significant structural and chemical changes during fruit maturation, including lignification and the incorporation of hydrolyzable tannins, which enhance its rigidity and protective properties[28]. We summarized the geometrical parameters extracted by natural fresh and dried samples, such as height and diameters, in Fig. 4 and Table 1.  We found that the total height ($H_t$) of natural samples ranged from 19.4±1.7 mm for dried mature fruits to 25.2±2.3 mm for fresh fruits. In addition, we compared the weights of dried and fresh samples (Table 2), finding that fresh samples are approximately 6.3 times heavier than dried samples due to the differences in the internal structures.

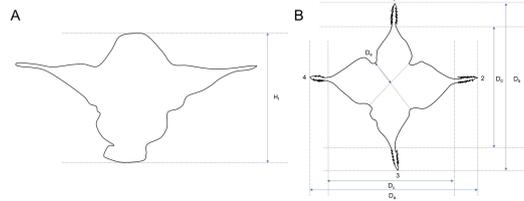

Fig. 4. Sketch of natural fruits of *T. natans*: (A) Front view, (B) Top view.

Table 1. Main geometrical parameters of natural samples of *T. natans*. These parameters include: $H_t$ (Total *T. natans* height), $D_a$ (Distance between the ends of 2 and 4 spines), $D_b$ (Distance between the ends of 1 and 3 spines), $D_c$ (Distance between the ends of 2 and 4 corners), $D_d$ (Distance between the ends of 1 and 3 corners), $D_e$ (Radius of the center section).

|  | Dried samples | | | Fresh samples | | |
| --- | --- | --- | --- | --- | --- | --- |
|  | Mean | S.D. | n | Mean | S.D. | n |
| $H_t$ [mm] | 19.4 | 1.7 | 4 | 25.2 | 2.3 | 6 |
| $D_a$ [mm] | 36.8 | 1.5 | 4 | 36.7 | 2.4 | 6 |
| $D_b$ [mm] | 38.9 | 1.6 | 4 | 40.1 | 1.2 | 6 |
| $D_c$ [mm] | 26.3 | 2.3 | 4 | 27.0 | 2.1 | 6 |
| $D_d$ [mm] | 27.0 | 1.8 | 4 | 29.4 | 1.3 | 6 |
| $D_e$ [mm] | 11.8 | 1.2 | 4 | 14.0 | 1.2 | 6 |

Table 2. Weight data of natural samples of *T. natans*.

|  | Weight [g] | | |
| --- | --- | --- | --- |
|  | Mean | S.D. | n |
| Dried samples | 0.7 | 0.1 | 4 |
| Fresh samples | 4.4 | 0.8 | 6 |

To create a biomimetic design of the fruits of *T. natans*, we performed X-Ray Micro-Computed Topography (microCT) experiments on dried fruits of *T. natans*. Through MicroCT, we obtained high-resolution 2D slice images of the shell structures of the dried mature fruits (Fig. 5A-D). We observed the anatomy of the fruit's endocarp (i.e., outer shell) in natural dried fruit of *T. natans*, which reveals a terminal pore (Fig. 5D). The dried fruit appears empty inside (Fig. 5D), in contrast to the fresh samples, which contained a living seed[21]. The anatomy of the outer shell of *T. natans*, as observed in our microCT investigations, is of great interest for artificial translation into a machine body. In fact, in nature, the fruits are designed to facilitate their transportation and spread across water bodies, often being carried by water flow[21].

After post-processing the 2D slices with 3D reconstruction softwares (please refer to Section 2.1, Materials and Methods, for more details), we obtained biomimetic 3D CAD models of *T. natans* fruits (Fig. 5E-K). Specifically, we obtained the CAD models for the machine body in two configurations, including a hollow and a solid machine body. The main geometrical parameters of the biomimetic CAD model are reported in Table 3. The CAD model of the machine body is scaled down 1:2 respect to the natural samples (please refer to dried samples in Table 1 and Fig. 4).

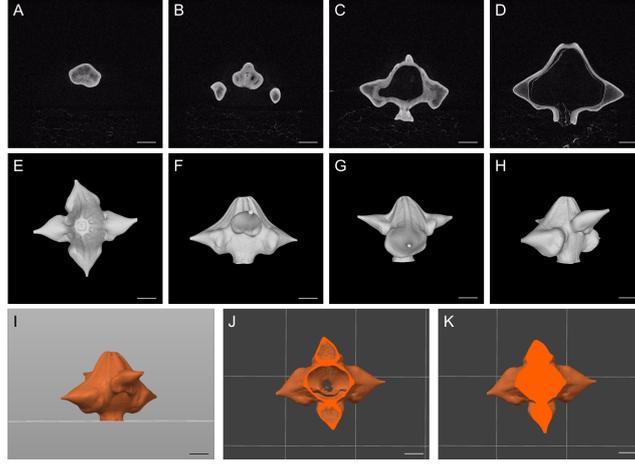

Fig. 5. Design of the bioinspired aquatic machine. (A-D) 2D slice images of natural *T. natans* fruits by MicroCT scanning. Scale bar, 5 mm. (E-H) 3D reconstruction of the natural *T. natans* fruits to obtain a 3D biomimetic CAD model. Scale bar, 5 mm. (I-K) CAD model of the natural *T. natans* fruits and sliced preview in PrusaSlicer of (J) hollow and (K) solid biomimetic models. Scale bar, 5 mm.

Table 3. Main geometrical parameters of the biomimetic CAD model for the machine body.

|  | Biomimetic design cad model |
|---|---|
| Total height, $H_t$ [mm] | 9.7 |
| Distance c, $D_c$ [mm] | 14.2 |
| Distance d, $D_d$ [mm] | 15.4 |

### 3.2 Bioprinting results of machine body

We successfully microfabricate the bioinspired aquatic machines mimicking the fruits of *T. natans* using a light-based Lumen X bioprinter and a photosensitive PEGDA X hydrogel (Fig. 6A-C). Compared to other 3D printing methods, photon-based bioprinting enables the creation of complex biomimetic structures by polymerizing bioinks containing cells, hydrogels, or other biomaterials[29], making it particularly suitable for the preparation of miniature machines. We microfabricate at high resolution the machine's body in two configurations: a hollow body (Fig. 6D-G) and a solid body (Fig. 6H-K). Both samples were observed to be completely intact without obvious defects or collapse, maintaining the shape integrity of the 3D CAD model of the natural *T. natans* (Fig. 6D-K). Compared to other soft robots that mimic plants[30, 31], our bionic machine accurately reproduces the 3D structure of the natural *T. natans* through Mi-croCT technology, realizing a higher degree of fidelity in the bioinspired design. Additionally, the printing results of the biomimetic *T. natans* samples were further examined using optical microscopy. The layer-by-layer stacking pattern, with no noticeable gaps, was clearly visible on the surface of both the hollow and solid biomimetic *T. natans* samples (Fig. 6L-O). Overall, the use of DLP bioprinting for the microfabrication of biomimetic machines, scaled 1:2 relative to the natural model, produced excellent results in terms of resolution and reproducibility (Fig. 6).

The geometric parameters of the hollow and solid biomimetic samples were compared immediately after printing and development (i.e., referred as 'Day 0') and after 7 days from the printing and development (i.e., referred as 'Day 7') (Table 4). In particular, the height of the bioprinted hollow and solid samples at day 0 is 9.65±0.19 mm and 9.61± 0.16 mm, respectively. The biomimetic structures show four corners, in which two corners have a slightly lower distance than the other two (Table 4). The corners show an end portion with a thin tip up to ~ 189 μm. After 7 days, both hollow and solid samples showed a similar swelling behavior, with an increase of the dimensions up to about 13 % (Table 4). In addition, we measured the weight of the hollow and solid biomimetic samples (Table 5). The results demonstrated high reproducibility between the replicates (both hollow and solid), with the solid samples exhibiting a weight approximately twice as high as the hollow ones (i.e., ~0.224 g, hollow samples; ~ 0.442 g, solid samples).

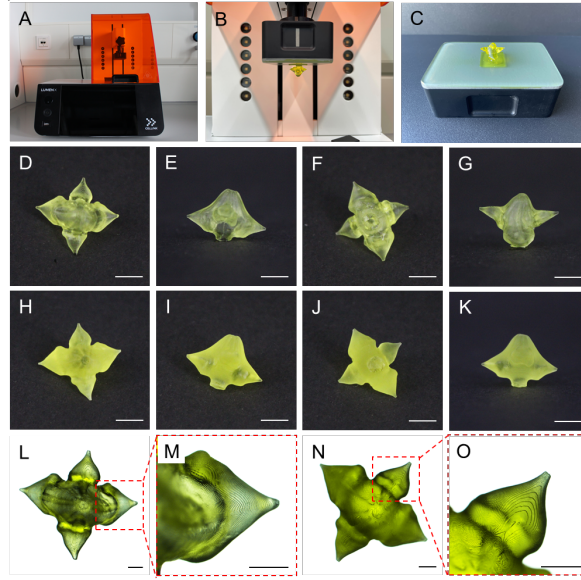

Fig. 6. Fabrication of bioinspired aquatic machine. (A-C) Biomimetic *T. natans* samples printed using LumenX. (D-G) Hollow biomimetic *T. natans* samples. Scale bar, 5 mm. (H-K) Solid biomimetic *T. natans* samples. Scale bar, 5 mm. (L-M) Hollow biomimetic *T. natans* samples and partial zoom of the optical microscopy images. Scale bar, 2 mm. (N-O) Solid biomimetic *T. natans* samples and partial zoom of the optical microscopy images. Scale bar, 2 mm.

Table 4. Main geometrical parameters of printed samples.

|  | Hollow samples | | | | | | Solid samples | | | | | |
|---|---|---|---|---|---|---|---|---|---|---|---|---|
|  | Day 0 | | | Day 7 | | | Day 0 | | | Day 7 | | |
|  | Mean | S.D. | n | Mean | S.D. | n | Mean | S.D. | n | Mean | S.D. | n |
| Total height, $H_t$ [mm] | 9.65 | 0.19 | 3 | 10.83 | 0.26 | 3 | 9.61 | 0.16 | 3 | 10.88 | 0.30 | 3 |
| Distance c, $D_c$ [mm] | 13.22 | 0.03 | 3 | 15.07 | 0.10 | 3 | 13.31 | 0.13 | 3 | 15.08 | 0.25 | 3 |
| Distance d, $D_d$ [mm] | 14.43 | 0.05 | 3 | 16.28 | 0.06 | 3 | 14.71 | 0.32 | 3 | 16.32 | 0.33 | 3 |

Table 5. Weight data of printed samples.

|  | Weight [g] | | |
|---|---|---|---|
|  | Mean | S.D. | n |
| Hollow samples | 0.224 | 0.002 | 3 |
| Solid samples | 0.442 | 0.013 | 3 |

### 3.3  Mechanical property analysis

After the microfabrication, we conducted compression tests using a universal tensile testing machine to investigate the mechanical properties under compressive loading of the bioprinted materials. Firstly, three standard cube samples (5x5 mm) were printed with the same printing parameters used for the biomimetic samples, and tested to investigate the mechanical properties of the material. The cube samples were placed on the machine and compressed until breakage to obtain the stress-strain curves (Fig. 7A). From the experiment, we extracted the average compression breaking strain (33.6 ± 0.8 %), the average compression breaking strength (8.0±0.3 MPa), and the average Young's modulus (15.4±0.4 MPa) (Table 6). Our results confirm that PEGDA X has not only showed high resolution and reproducibility, but has also excellent mechanical strength and mechanical stability. Furthermore, we also conducted preliminarily compression tests on the solid biomimetic samples, obtaining the force-displacement curves reaching a force up to 7.9 N (Fig. 7B).

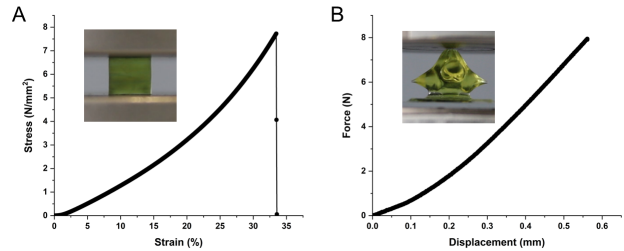

Fig. 7. Main results of the compression test. (A) Stress-strain curve of a test cube. (B) Force-displacement curve of solid biomimetic samples.

Table 6. Main mechanical parameters of the bioprinted materials.

|  | Bioprinted test cube (PEGDA X) | | |
| --- | --- | --- | --- |
|  | Mean | S.D. | n |
| Young modulus [MPa] | 15.4 | 0.4 | 3 |
| Compression Breaking strain [%] | 33.6 | 0.8 | 3 |
| Compression Breaking strength [MPa] | 8.0 | 0.3 | 3 |

### 3.4 *Proof-of-concept* demonstration for machine dispersal mechanism in water

We performed a *proof-of-concept* investigation of the dispersal mechanisms of the bioprinted aquatic machines in water using a dedicated setup with a flow channel (Fig. 8). The natural fruits of *T. natans* exhibited two distinct behaviors in water: 1) fresh fruits (containing living seeds) sink to the bottom and can be transported by the water flow[21], while 2) mature dried fruits (i.e., that are empty inside) remain on the water surface and can also be transported by the water flow (Fig. 8A). In order to create biomimetic aquatic machines able to work above or inside (under-)water environments, we bioprinted the machines in hollow and solid configurations (as showed in Fig. 6D-H). Similarly to *T. natans* fruits, our machines can sink to the bottom or remain above water surface (Fig. 8B). Specifically, the results showed that the hollow biomimetic samples were able to float on the water surface and can be transported with the water flow (Fig. 8C). In contrast, solid biomimetic samples were able to sink to the bottom of the water like natural fresh *T. natans* and exhibited relatively slow motility with the same velocity of water flow (Fig. 8D). Based on the above results, we can conclude that the structural design and material density significantly affect the underwater motion characteristics of the bioinspired aquatic machine. Furthermore, reasonable optimization of its structure and density could be used in the future to regulate the machine's (under-)water behavior.

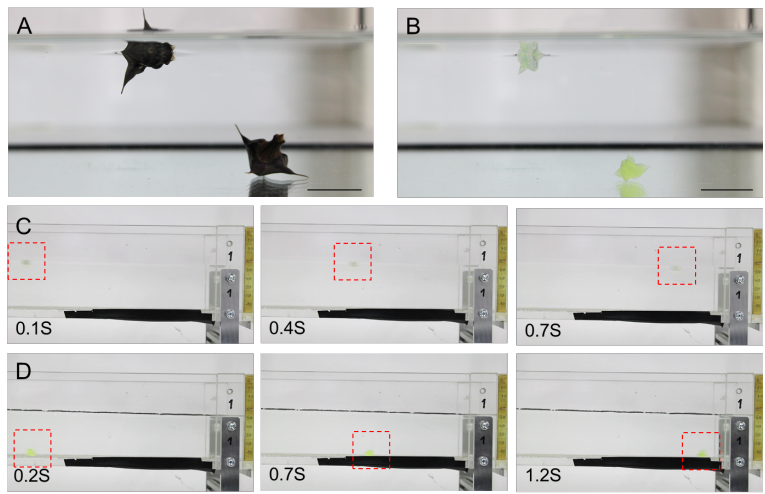

Contribution Title (shortened if too long)        11Fig. 8. Proof-of-concept for machine dispersal mechanisms in water. (A-B) Stationary state of (A) natural fruits of *T. natans* and (B) biomimetic aquatic machines in water. Scale bar, 2 cm. (C) Movement of hollow biomimetic samples in water. (D) Movement of solid biomimetic samples in water.

## 4     Conclusion and future perspectives

In conclusion, this study presents the first *state-of-the art* bioinspired aquatic machine that mimics the unique structure and dispersal mechanisms of the water caltrop (*Trapa natans* L.). We successfully fabricated at high resolution the biomimetic machines in two configurations (hollow body and solid body) using DLP 3D bioprinting, which demonstrated to be an effective fabrication method to mimic the structures of the natural fruits of *T. natans*. In addition, we preliminarily characterize the mechanical properties of the bioprinted materials, and we compared the motion behavior in water of the bioprinted hollow and solid biomimetic machines in a flow channel under a certain flow condition. As future perspectives, the geometrical details of the machines and the mechanical data of the hydrogels will be used as input parameters for a mathematical model that will help investigating the behavior of biomimetic aquatic machines under different flow conditions in underwater environments. Moreover, further improvements of this work will focus on mimicking not only the machine's body but also the spines of the natural fruits of *T. natans*. In fact, the spines of *T. natans* play a key role in fruits dispersal by providing attachment to the underwater soil and/or the surrounding environments, and will be integrated in the proposed machine's body. In addition, we are considering integrating multifunctional sensors in this robot to further enhance its potential for applications in marine environment monitoring. Overall, these results represent a starting point for developing multifunctional, self-dispersing aquatic machines inspired by plants, which could have applications in underwater environmental monitoring and seeding, providing a new direction for the development of bioinspired aquatic miniaturized machines.

**Acknowledgments.** We would like to thanks Jessika Huss and Sebastian J. Antreich from Institute of Biophysics (University of Natural Resources and Life Sciences (BOKU), 1190 Vienna, Austria) for providing the dry samples of *Trapa natans* fruits. We would also like to thank David Schultz for his technical assistance. Special thanks to Madeleine Tomasovic Savigny from Cellink for the fruitful discussions on bioprinting. This study was funded by the Deutsche Forschungsgemeinschaft (DFG, German Research Foundation) under Germany's Excellence Strategy - EXC-2193/1-390951807.**Disclosure of Interests.** Y.L. and I.F plan to have a pending patent application related to this project.